\documentclass[a4paper,11pt]{article}
\usepackage{jheppub}
\usepackage[T1]{fontenc}
\usepackage{amsfonts}
\usepackage{amsmath}
\usepackage{amssymb,epsf}
\usepackage{latexsym}
\usepackage{graphicx,epsfig}
\usepackage{amssymb}
\usepackage{subfigure}

\title{Mimetic Black Strings}
\author{Ahmad Sheykhi}
\affiliation{Physics Department and Biruni Observatory, Shiraz
University, Shiraz 71454, Iran} \affiliation{ Max-Planck-Institute
for Gravitational Physics (Albert-Einstein-Institute), 14476
Potsdam, Germany} \emailAdd{asheykhi@shirazu.ac.ir}

 \abstract{We present two new classes of black string solutions in the
context of mimetic gravity. The horizon topology of these
solutions can be either a flat $T^2$ torus with topology $S^1
\times S^1$, or a standard cylindrical model with topology
$R\times S^1$. The first class describes uncharged rotating black
string which its asymptotic behavior is a quotient of anti-de
Sitter (AdS) space, while the second class represents
asymptotically AdS charged rotating black string. We study the
casual structure and physical properties of these spacetimes and
calculate, the entropy, electric charge, mass and angular momentum
per unit length of rotating black strings.}

\begin{document}
\maketitle \flushbottom

\section{Introduction }
The idea of the mimetic gravity was proposed \cite{Mim1}, in order
to shed the light on the problem of dark matter. Roughly speaking,
isolating the conformal degree of freedom of the gravitational
field in a covariant way, it was shown that the conformal degree
of freedom can be dynamical even in the absence of matter. As a
result, the extra scalar longitudinal degree of freedom of the
gravitation field, can be interpreted as the energy density of the
mimetic field which scales in term of the scalae factor $a$, as
$a^{-3}$, which resembles the contribution of pressureless dust,
without needing to particle dark matter \cite{Mim1}. In the past
few years, mimetic theory of gravity has arisen a considerable
attentions both in the cosmological setup as well as in stationary
background. Cosmological dynamics of mimetic gravity, with various
form of the potential for the mimetic field, has been investigated
\cite{MimCos,Dutta}. It has been confirmed that this scenario can
describe the thermal history of the universe, namely the
successive sequence of radiation, matter, and dark-energy eras
\cite{Dutta}. Besides, at late times, mimetic cosmology is capable
to alleviate the cosmic coincidence problem and can be brought
into good agreement with observational data \cite{Dutta}. It has
been realized that with modification to the action of mimetic
gravity, this model can address the singularities in contracting
Friedmann and Kasner universes and yields a universe with limiting
curvature and regular bounce \cite{MimCos1}. Considering the
coupling between mimetic scalar field with a vector field, an
anisotropic model of mimetic cosmology was explored in \cite{Sep}.
While inflationary model in the context of mimetic $f(G)$ gravity
has been explored \cite{Zh}, a unified description of dark energy
and dark matter in the context of mimetic theory was suggested in
\cite{Mat}. Other studies on the mimetic cosmology can be found in
\cite{Odin0,Odin1,Odin2,MimCos2,Gorj1,Gorj2,Gorj3,Gorj4,Leb,Cham2}.

Parallel to the mimetic cosmology, the studies on static and
stationary solutions of mimetic gravity started and have also
being continued. Static and spherically symmetric black hole
solutions of mimetic gravity have been explored in
\cite{Myr1,Myr2}. The studies have also been generalized to
modified Gauss-Bonnet  mimetic gravity \cite{OdinGB} as well as
$f(R)$ mimetic theory \cite{OdinP, Odinfr1,Odinfr2,Oik}. In
particular, it has been realized that, by suitably chosen mimetic
potential for a given $f(R)$ gravity, this model admits an
inflationary cosmology compatible with Planck observations
\cite{OdinP}. It was also observed that by modifying the action of
mimetic gravity, a regular black hole can be exist without
singularity inside the Schwarzschild spacetime \cite{Cham3}.
Topological black holes in mimetic gravity in the presence of
constant potential for the mimetic field were recently obtained in
\cite{ShSa}. By gluing an exterior static solution to a
time-dependent anisotropic interior solution, the authors of
\cite{Gorji2} constructed mimetic black hole spacetime. They also
revealed that these two solutions match continuously on the event
horizon \cite{Gorji2}. The authors of \cite{Nash1,Nash2}
considered static and rotating solutions of mimetic gravity with
flat or cylindrical horizon in the presence of cosmological
constant. However, the obtained solution in \cite{Nash1,Nash2} are
indeed nothing but the well-known black brane/string solutions of
Einstein gravity. The reason is that they have simply assumed
$\textbf{g}_{tt}=-\textbf{g}^{rr}$, and arrived at a special
solution of the mimetic field equations. In order to reflect the
impact of the extra degree of freedom of the gravitational field
equations, encoded in the mimetic scalar field, on the metric
functions, one need to consider the more general spacetime with
$\textbf{g}_{tt}\neq-\textbf{g}^{rr}$
\cite{Myr1,Myr2,ShSa,Gorji2,Ch,Nash3,Der}. Other studies on
mimetic gravity can be carried out in
\cite{Bar,Gol,Lim,Cha,Mal,jibril}.

One of the main question concerning mimetic theory, is that
whether this theory is capable to reproduce the inferred flat
rotation curves of galaxies. It was shown that adding a
non-minimal coupling between matter and mimetic field yields the
appearance of an extra force and the resulting equations of motion
in the weak field limit provides a Modified Newtonian Dynamics
(MOND)-like \cite{MimMOND} theory which may alleviate the flat
rotation curves of spiral galaxies without needing to particle
dark matter. Also, constructing the potential associated with the
mimetic field, by applying an inverse approach, for a general
static spherically symmetric spacetime, it has been shown that the
corrections to the Schwarzschild spacetime could provide an
explanation for the inferred flat rotation curves of spiral
galaxies within the mimetic gravity framework \cite{Myr2}. More
recently, it has been realized \cite{ShSa} that one is capable to
reproduce the flat galactic rotation curves, in the context of
mimetic gravity, without neither modifying the action
\cite{MimMOND} nor taking  into account a complicated potential
for the mimetic field \cite{Myr2}. Indeed, the original theory of
mimetic gravity \cite{Mim1} was proposed in such a way that it
naturally mimics dark matter as a geometrical effect. It was
argued that the origin of the responsible term for recovering the
flat galactic rotation curves in a static spacetime, is exactly
the one which mimics \textit{dark matter} in the cosmological
setup \cite{ShSa}.

In this paper, we are going to construct the general rotating
black string solutions in the context of mimetic gravity. The
rotating solutions of the Einstein equation with a negative
cosmological constant with cylindrical and toroidal horizons have
been investigated in \cite{Lem1}. The extension to include the
Maxwell field has been done and the static and rotating
electrically charged black string have been considered in
\cite{Lem2,Awad,mhd,mhdk}. When the gauge field is in the form of
nonlinear electrodynamics, rotating black string solutions have
been explored in \cite{H,HS}. The studies on the rotating black
strings were also generalized to dilaton gravity \cite{mhdF,She1},
holographic superfluids \cite{Betti} and $f(R)$ gravity
\cite{She2}.

This paper is structured as follows. In Sec. \ref{Field}, we
introduce the action and basic field equations. In Sec.
\ref{Uncha}, we analyze rotating black string in the context of
mimetic gravity. In Sec. \ref{Cha}, we generalize our study to the
case of charged rotating mimetic black string. We finish with
conclusion and discussion in Sec. \ref{Con}.

\section{Field equations} \label{Field}
The action of mimetic gravity in the presence of mimetic potential
is \cite{MimCos}
\begin{eqnarray}\label{Act}
&&S=\int{d^4x\sqrt{-{g}}\left[\mathcal{R}+{\lambda}(\partial
^{\mu} \phi
\partial _{\mu}\phi-\epsilon)-V(\phi)+
\mathcal{L}_{m}\right]},
\end{eqnarray}
where $\mathcal{R}$ is the Ricci scalar curvature, $\lambda$ is
the Lagrange multiplier, and $V=V(\phi)$ is the potential for the
mimetic scalar field $\phi$. Here $\mathcal{L}_{m}=-F_{\mu \nu }
F^{\mu \nu }$ is the Lagrangian of the Maxwell field, where
$F_{\mu \nu }=\partial _{\mu }A_{\nu }-\partial _{\nu }A_{\mu }$
is the electromagnetic field tensor and $A_{\mu }$ is the gauge
potential. Besides, $\epsilon=\pm1$ depends on the spacelike or
timelike nature of the vector field $\partial _{\mu} \phi$.
$\epsilon=+1$ corresponds to spacelike vector field $\partial
_{\mu} \phi$, while $\epsilon=-1$ indicates timelike $\partial
_{\mu} \phi$. Through this paper we set $8\pi G_N=1$ and work in
the metric signature $(-,+,+,+)$. From the variation of the above
action with respect to $\lambda$, one immediately finds
\begin{equation}\label{cond}
g^{\mu \nu}\partial _{\mu} \phi \partial _{\nu} \phi=\epsilon.
\end{equation}
This constraint imposes strong restriction on the functional form
of $\phi$ which in general depends on the metric line elements of
the underlying theory. This condition can also be understood by
assuming the physical metric $g_{\mu\nu}$ to be a function of two
variable; namely an auxiliary metric $\tilde{g}_{\mu \nu}$ plus a
scalar field $\phi$ and redefine the metric as $g_{\mu \nu}=
\epsilon (\tilde{g}^{\alpha \beta}
\partial _{\alpha} \phi
\partial _{\beta} \phi)\tilde{g}_{\mu \nu}$ \cite{Mim1}. Then, one can
immediately recovers (\ref{cond}). It was argued that considering
this non-invertible conformal/disformal transformation, increases
the number of degrees of freedom in such a way that in addition to
two transverse degrees of freedom describing gravitons, the
gravitational field acquires an extra dynamical longitudinal mode
\cite{Der,Fir}.

The equations of motion can be derived by varying the action with
respect to the metric $g_{\mu \nu}$, the scalar field $\phi$, and
the gauge field $A_{\mu}$.  The result is
\begin{eqnarray}\label{FE1}
{G}_{\mu\nu}&=& \lambda \partial _{\mu} \phi \partial _{\nu}
\phi-\frac{1}{2}g_{\mu \nu }V(\phi)+T_{\mu \nu },
\end{eqnarray}
\begin{equation}\label{FE2}
\nabla ^{\mu}(\lambda \partial _{\mu} \phi )=-\frac{1}{2}\frac{d V
(\phi)}{d \phi},
\end{equation}
\begin{equation}
\partial _{\mu }\left( \sqrt{-g} F^{\mu \nu }\right)
=0, \label{FE3}
\end{equation}%
where the energy momentum tensor of the Maxwell field is given by
\begin{equation}\label{Tem}
T_{\mu \nu }=2 F_{\mu \gamma } F_{ \nu }^{\ \gamma}-\frac{1}{2}
g_{\mu \nu } F_{\alpha \beta } F^{\alpha \beta },
\end{equation}
Now, we take the trace of Eq. (\ref{FE1}), combining with Eq.
(\ref{cond}), yields
\begin{eqnarray}\label{lambda}
\lambda=\epsilon\left(G-T+2V\right),
\end{eqnarray}
where $G=G^{\mu}_{\mu}$ and $T=T^{\mu}_{\mu}$ are, respectively,
the trace of the Einstein tensor and energy momentum tensor.
Inserting $\lambda$ in the field Eqs. (\ref {FE1}) and (\ref
{FE2}), we arrive at
\begin{eqnarray}\label{FEE1}
{G}_{\mu\nu}&=& \epsilon(G-T+2V) \partial _{\mu} \phi \partial
_{\nu} \phi-\frac{1}{2}g_{\mu \nu }V(\phi)+T_{\mu \nu },
\end{eqnarray}
\begin{equation}\label{FEE2}
\nabla ^{\mu}[(G-T+2V)\partial _{\mu}
\phi]=-\frac{\epsilon}{2}\frac{d V (\phi)}{d \phi}.
\end{equation}
Our aim here is to construct rotating black string solutions of
the above field equations and investigate their properties. The
metric of four-dimensional rotating solution with cylindrical or
toroidal horizons can be written as \cite{Lem1,Lem2,Awad}
\begin{eqnarray}\label{metric1}
ds^{2} &=&-f(r) g^2(r)\left( \Xi dt-a d\varphi \right)
^{2}+\frac{dr^{2}}{f(r)}+ \frac{r^2}{l^4}\left(a dt-\Xi l^2
d\varphi \right) ^{2}
+\frac{r^{2}}{l^{2}}dz^{2},   \\
\Xi ^{2} &=&1+\frac{a^{2}}{l^{2}}, \nonumber
\end{eqnarray}
where the constants $a$ and $l$ have dimensions of length and as
we will see later, $a$ is the rotation parameter and $l$ can be
interpreted as the AdS radius, in case of constant potential.
Compared to the line element of the rotating black string of
Einstein gravity \cite{Awad}, here we introduce a new function
$g(r)$ in the metric due to the extra degree of freedom of the
gravitational field, encoded by mimetic field. By solving the
field equations, the unknown functions $f(r)$ and $g(r)$ would be
determined. In metric (\ref{metric1}) the ranges of the time and
radial coordinates are $-\infty<t<\infty$, $0\leq r< \infty$. We
are going to consider solutions in mimetic gravity with
cylindrical symmetry. This implies that the spacetimes admit a
commutative two dimensional Lie group $G_2$ of isometries
\cite{Lem2}. The topology of the two dimensional surface,
$t$=constant and $r$ =constant, generated by $G_2$ can be (i) the
flat $T^2$ torus model with topology $S^1 \times S^1$ $[\rm{i.e.}\
G_2=U(1)\times U(1)]$, and $0\leq \varphi<2\pi$, $0\leq z<2\pi l$,
(ii) the standard cylindrical model with topology $R\times S^1$
$[\rm{i.e.}\ G_2=R\times U(1)]$, and $0\leq \varphi<2\pi$,
$-\infty< z<\infty$,  and (iii) the infinite plane $R^2$ with
$-\infty< \varphi<\infty$ and $-\infty< z<\infty$ (this planar
solution does not rotate). Hereafter, we shall consider the
topology (i) and (ii).

In the coordinate system with metric (\ref{metric1}), the general
solution to the constraint equation (\ref{cond}) become
\begin{equation}\label{phi}
f(r) \phi'^{2}=\epsilon,   \   \rightarrow   \   \phi(r)=\pm
\int{\frac{dr}{ \sqrt{|\epsilon f(r)|}}}+\rm const,
\end{equation}
where without loss of generality one may choose $+$ sign of
integral. The scalar field is always constrained by Eq.
(\ref{cond}) or, equality in our case, by Eq. (\ref{phi}).
Therefore, we have no new degrees of freedom for the scalar field
$\phi$. Indeed, this constrained scalar field is not dynamical by
itself, but induces mimetic dark matter in Einstein theory making
the longitudinal mode of gravity dynamical \cite{Mim1}.  Let us
remind that in the cosmological setup the vector field $\partial
^{\mu} \phi$ is always timelike because it plays the role of the
four-velocity $u^{\mu}$ of the perfect fluid, where $\phi$ is the
velocity potential \cite{Mim1}. Indeed, in the background of
Friedmann-Robertson-Walker universe the general solution of the
constraint equation (\ref{cond}) is $\phi=\pm t+A$, where $A$ is
an integration constant, and without loss of generality one can
identify $\phi=t$ \cite{MimCos}. However, in the context of black
hole/string, where, by definition, the spacetime has an event
horizon, the vector field $\partial _{\mu} \phi$ changes from a
spacelike vector to a timelike vector over the horizon. However,
since, the scalar field $\phi$ is not dynamical by itself, one may
not worry about this behvaiour and can still search for black
hole/string solutions with real $f(r)$.

Considering the metric (\ref{metric1}), the gauge potential takes
the form
\begin{equation}\label{A}
A_{\mu}=h(r) \left(\Xi\delta_\mu^{0}-a\delta_\mu^{\phi}\right).
\end{equation}
With this gauge potential at hand, the nonvanishing components of
the electromagnetic tensor become
\begin{equation}\label{FtrFpr}
F_{tr}=\Xi h'(r),  \   \   \  \   \   F_{\varphi r} =-a
h'(r)=-\frac{a}{\Xi}F_{tr}.
\end{equation}
Inserting the gauge potential (\ref{A}) into Eq. (\ref{FE3}), one
finds that the non-vanishing components of the Maxwell equation
are just the ($t$) and $(\varphi)$ components, which both of them
yield the following equation
 \begin{equation}\label{hr}
rh''(r)g(r)-rh'(r)g'(r)+2h'(r)g(r)=0,
\end{equation}
where by prime we denote $r$ derivative. The above equation admits
the following unique solution
\begin{equation}\label{Ftr}
h'(r)=\frac{q}{r^2}g(r),
\end{equation}
where we denote the constant of integration by $q$, since it
should be related to the electric charge of the black hole. Hence
from (\ref{FtrFpr}), we obtain
\begin{equation}\label{FtrFpr2}
F_{tr}=\Xi \frac{q}{r^2}g(r) ,  \    \  \   \   F_{\varphi r} =-a
\frac{q}{r^2}g(r).
\end{equation}
Inserting metric (\ref{metric1}), condition (\ref{phi}), and the
electromagnetic fields (\ref{FtrFpr2}) into the field equations
(\ref{FEE1}), we arrive at the following equations for the
non-vanishing components of the gravitational field equations
\begin{eqnarray}\label{tt}
&&g l^2 \Xi^2  \left(2r^3 f^{\prime}+2f {r}^{2}+V \left(
\phi\right) {r}^{4}+2{q}^{2}\right)+a^2 g\left(-2r^3 f'-r^4f''-
r^4 V(\phi)+2q^2\right)\nonumber\\&&+a^2\left(-2r^4f
g''-2r^3fg'-3r^4f'g'\right)=0,
\end{eqnarray}
\begin{eqnarray}\label{rr}
&&3\, {r}^{3} g f^{\prime}+2\, {r}^{3}f g^{\prime} +{r}^{2}g
  f +3\, {r}^{4} f^{\prime}g^{\prime}+2\, {r}^{4} f  g'' +{r}^{4} g  f''
  +\frac{3}{2}{r}^{4} g
 V \left( \phi \right) -{q}^{2}g =0,
\end{eqnarray}
\begin{eqnarray}\label{pp}
&&\Xi^2 l^2 \left[2\, {r}^{3} (f  g)' +3 {r}^{4}f' g'+2 {r}^{4} f
g''+{r}^{4} g f'' + {r}^{4} \,g
 V \left( \phi \right) -2\,{q}^{2}g\right]\nonumber
\\
&& -a^2 g\left(2r^3 f'+2r^2f +{r^4}V(\phi)+2q^2\right)=0,
\end{eqnarray}
\begin{eqnarray}\label{zz}
&& 2\, {r}^{3} (f  g)' +3 {r}^{4}f' g'+2 {r}^{4} f
 g''+{r}^{4} g f'' + {r}^{4} \,g
 V \left( \phi \right) -2\,{q}^{2}g =0,
\end{eqnarray}
\begin{eqnarray}\label{pt-tp}
&& \Xi a \left(3 r^4 f' g'-2gf r^2-4q^2g +2r^3f g'+r^4g f''+2r^4 f
g'' \right)=0.
\end{eqnarray}
Our aim here is to solve the above field equations and obtain the
unknown functions $f(r)$ and $g(r)$. Clearly, our solutions should
also satisfy Eq. (\ref{FEE2}) for the scalar field. During this
paper we consider the case for which the potential is constant,
namely, $V(\phi)=V_{0}=-2\Lambda$, where for later convenience we
assume this constant to be positive definite, i.e.
$\Lambda=3/l^2>0$.

Multiplying Eq. (\ref{zz}) by factor $\Xi^2 l^2$ and subtracting
the resulting equation from (\ref{pp}), we arrive at
\begin{eqnarray}\label{ppzz}
r^3 f'+r^2f +\frac{r^4}{2}V(\phi)+q^2=0.
\end{eqnarray}
Substituting $V(\phi)=-2\Lambda$ in Eq. (\ref{ppzz}), and solving
for $f(r)$, we find
\begin{eqnarray}\label{fr1}
f(r)&=&-\frac{m}{r}+\frac{q^2}{r^2}+\frac{\Lambda}{3}r^2,
\end{eqnarray}
where $m$ and $q$ are integration constants which are,
respectively, related to the mass and charge of the black string.
Inserting solution (\ref{fr1}) into Eq. (\ref{rr}), we find the
following equation for the metric function $g(r)$:
\begin{eqnarray}\label{eqg}
(2\Lambda r^5-6 mr^2+6q^2 r )g''+(8\Lambda r^4+3mr-12 q^2)g'=0.
\nonumber\\
\end{eqnarray}
Solving yields
\begin{eqnarray}\label{gr0}
g(r)=c_{1}+c_2 \int{\frac{r^2 dr}{(|\Lambda r^4 -3m
r+3q^2|)^{3/2}}},
\end{eqnarray}
where $c_1$ and $c_2$ are the constants of integration. In the
limiting case where $c_2=0$, our solution is nothing but the
corresponding one of Einstein gravity which is indeed the solution
obtained in \cite{Nash1}. Without loss of generality we can set
$c_1=1$. Indeed, the effects of $c_1$ in function (\ref{gr0}) can
be absorbed by rescaling $t\rightarrow t/c_1$ and $\varphi
\rightarrow \varphi/c_1$ in the line element (\ref{metric1}).
Thus, we rewrite the general solution as
\begin{eqnarray}\label{gr1}
g(r)=1+b\int{\frac{r^2 dr}{(|\Lambda r^4 -3m r+3q^2|)^{3/2}}},
\end{eqnarray}
where $b$ is a constant which reflects the imprint of the mimetic
field in the solutions. One can easily check that the obtained
solutions (\ref{fr1}) and (\ref{gr1}) are fully satisfied the
field equations (\ref{FEE2}) and (\ref{tt})-(\ref{pt-tp}). Since
the integral in the function (\ref{gr1}) cannot be calculated
easily, we consider in the following two sections the uncharged
and charged rotating black strings, separately.
\section{Uncharged mimetic black string} \label{Uncha}
In this case ($q=0$), the field Eq. (\ref{eqg}) reduces to
\begin{eqnarray}\label{eqg2}
(2\Lambda r^5-6 mr^2)g''+(8\Lambda r^4+3mr)g'=0,
\end{eqnarray}
which admit the following solution
\begin{eqnarray}\label{gr2}
g(r)=1+\frac{b_0 \ r^{3/2}}{\sqrt{|\Lambda r^3 -3m}|},
\end{eqnarray}
where $b_{0}$ is a constant with dimension $[length]^{-1}$ which
incorporates the imprint of the mimetic field on the spacetime
metric. The horizon, by its definition, is a null hypersurface
$\mathcal {S}$ at constant $r=r_{+}$, determined by
$\textbf{g}^{rr}=0$ \cite{Townsend}. For the black string
spacetime under consideration, this implies
$\textbf{g}^{rr}=f(r_{+})=0$ which, using (\ref{fr1}) with $q=0$,
leads to $r_{+}=(3m/\Lambda)^{1/3}$.  The sign in the denominator
of (\ref{gr2}) depends on inner ($r<r_{+}$) or outer ($r>r_{+}$)
part of the solutions. For $r>r_{+}$,  $|\Lambda r^3 -3m|=\Lambda
r^3 -3m$, while for $r<r_{+}$, $|\Lambda r^3 -3m|=3m-\Lambda r^3$.
The spacetime is regular for whole range $0<r<\infty$ and the
divergency in $g(r)$ at $r=r_{+}$ is just a coordinate singularity
and all scalar invariants are finite in the whole range
$0<r<\infty$. Our solution is not static and the Killing vector
\begin{eqnarray}\label{chi}
\chi=\partial_{t}+\Omega_{+} \partial_{\varphi},
\end{eqnarray}
is the null generator of the event horizon where $\Omega_{+}$ is
the angular velocity of the event horizon. Since the Killing
vector (\ref{chi}) is normal to the null hypersurface $\mathcal S$
($r=r_{+}$), thus $\mathcal {S}$ is a Killing horizon of the
Killing vector field  \cite{Townsend}. By analytic continuation of
the metric we can obtain the surface gravity and angular velocity
associated with the horizon. The analytical continuation of the
Lorentzian metric by $t\rightarrow i\tau$ and $a\rightarrow ia$
yields the Euclidean section, whose regularity at $r = r_+$
requires that we should identify $\tau\sim \tau+\beta_+$ and
$\phi\sim\phi+i\beta _+\Omega_+$ where $\beta_+$ and $\Omega_+$
are the inverse Hawking temperature and the angular velocity of
the horizon \cite{Townsend}. It is a matter of calculation to show
that
\begin{eqnarray}\label{TOmega}
\beta _+ &=&\frac{4 \pi \Xi}{g(r_{+}) f'(r_{+})}, \\
\Omega_+&=&\frac{a}{\Xi l^2}.
\end{eqnarray}
Of course the temperature can also be calculated via surface
gravity. The general definition of surface gravity is \cite{Wald}
\begin{eqnarray}\label{TOmega}
\kappa=\sqrt{-\frac{1}{2} (\nabla ^{\mu} \chi^{\nu})(\nabla _{\mu}
\chi_{\nu}) },
\end{eqnarray}
where in our case, $\chi^{\nu}=(1,0,\Omega_+,0)$ is the Killing
vector (\ref{chi}) associated with the Killing horizon $\mathcal
{S}$. A simple calculation for the spacetime (\ref{metric1}) gives
$\kappa=g(r_{+}) f'(r_{+})/(2 \Xi)$.

One may also calculate the entropy associated with the horizon.
The entropy of the black string still obeys the so called area law
of the entropy which states that the entropy of the black hole is
a quarter of the event horizon area. It is a matter of calculation
to show that the entropy per unit length of the string is
\begin{equation}\label{ent}
{S}=\frac{r_{+}^2 \Xi}{4l}.
\end{equation}

In order to see the asymptotic behavior of the solutions, we
consider the non-rotating case ($a=0$), where the line elements of
the black string is given by
\begin{eqnarray}\label{metric2}
ds^{2} &=&-f(r) g^2(r)dt^2+\frac{dr^{2}}{f(r)}+r^2 d\varphi^2
+\frac{r^{2}}{l^{2}}dz^{2}.
\end{eqnarray}
Therefore, the $(tt)$ component of the static black string is
given by ${-\textbf{g}_{tt}(r)}=f(r) g^2(r)=B(r)$, where
\begin{eqnarray}\label{g00}
B(r)=\left(-\frac{m}{r}+\frac{\Lambda
r^2}{3}\right)\left[1+\frac{b_0 \ r^{3/2}}{\sqrt{|\Lambda r^3
-3m|}}\right]^2.
\end{eqnarray}
Expanding $g(r)$ and $B(r)$ for large values of $r$, we arrive at
\begin{eqnarray}\label{grexp1}
&&g(r)\approx 1+\frac{b_0}{\sqrt{\Lambda}}+\frac{3m b_0}{2\Lambda^{3/2}r^3}+O\left(\frac{1}{r^6}\right),  \\
&&B(r)\approx  \frac{r^2}{3}\left( \Lambda+b_0^2+2b_0
\sqrt{\Lambda}\right)-\frac{m}{r}
\left(1+\frac{b_0}{\sqrt{\Lambda}}\right)+O\left(\frac{1}{r^4}\right).
\label{gttexp1}
\end{eqnarray}
Thus as $r\rightarrow \infty$, the spacetime is a quotient of AdS
\footnote{For asymptotic AdS spacetime one expects
$f(r)=B(r)\approx \Lambda r^2/3$. In this paper, our notation is
such that we have defined $\Lambda=+3/l^2>0$, and therefore
$\Lambda>0$ corresponds to AdS space. The reason for this choice
is that the term $\sqrt{\Lambda}$ appears in our solutions, which
implies that it should be positive definite. } although  its
asymptotic behavior get modified due to the presence of the
mimetic field $\phi$ which its imprint incorporates in the metric
functions via the constant $b_0$.  Nevertheless, by redefining $
 (\Lambda+b_0^2+2b_0 \sqrt{\Lambda})\mapsto \Lambda '$, one can easily
realize that the solution given in (\ref{gttexp1}) is
asymptotically AdS.

It is also interesting to calculate the curvature scalars of this
spacetime. Expanding the Ricci scalar and the Kretschmann
invariant around $r=0$, we arrive at
\begin{eqnarray}
&&\lim_{r\rightarrow0}R =\frac{\rm const.} {\sqrt{r^{3}}}-4 \Lambda+b_0^2+\cdot\cdot\cdot,  \label{Rorigin} \\
&&\lim_{r\rightarrow 0}R_{\mu \nu \rho \sigma }R^{\mu \nu \rho
\sigma }=\frac{12 m^2}{r^6}-\frac{17 m b_0^2}{r^3}+
\cdot\cdot\cdot \nonumber
 \label{RRorigin}
\end{eqnarray}
Thus, both curvature invariants diverge as $r\rightarrow 0$, which
confirms that there is an essential singularity located at $r=0$.
On the other side, for the asymptotic region where $r\rightarrow
\infty$, the invariants of the spacetime are obtained as
\begin{eqnarray}
\lim_{r\rightarrow \infty }R &=&-4\Lambda ,  \label{Rinf} \\
\lim_{r\rightarrow \infty }R_{\mu \nu \rho \sigma }R^{\mu \nu \rho
\sigma } &=&\frac{8}{3}\Lambda^{2}, \label{RRinf}
\end{eqnarray}
It is also easy to show that Ricci and Kretschmann invariants have
finite values on the horizon $r_{+}$. Having a look on Eq.
(\ref{g00}), one may wonder that at the horizon, the function
$-\textbf{g}_{tt}(r)=B(r)$ diverge and the metric becomes
singular. However, this is indeed not the case. Simplifying the
metric function (\ref{g00}), it can be rewritten
\begin{eqnarray}\label{gg00}
B(r)=\frac{\left(\sqrt{\Lambda r^3 -3m}+b_0 r^{3/2}\right)^2}{3r}
\end{eqnarray}
which clearly has a finite value at $r_{+}$. Furthermore, from the
above expression we observe that $B(r)$ is well-defined for
$r>r_{+}$ and has a minimum value $B(r)\mid _{\rm min}=b_0^2
r_{+}^2/3$ at the horizon.

One may also look for the surface where the  redshift becomes
infinite. The infinite redshift surface is the surface where clock
intervals in flat space correspond to infinite time intervals on
the infinite redshift surface. More precisely, for an emitter $E$
and receiver $R$ with fixed spatial coordinates in a stationary
spacetime, the gravitational frequency shift of a photon is, quite
generally, given by \cite{Hob}
\begin{eqnarray}\label{IRS}
\frac{\nu_{R}}{\nu_{E}}=\left[
\frac{\textbf{g}_{tt}(A)}{\textbf{g}_{tt}(B)}\right]^{1/2},
\end{eqnarray}
where $A$ is the event at which the photon is emitted and $B$ the
event at which it is received. Thus, we see that if the photon is
emitted from a point with fixed spatial coordinates, then
$\nu_{R}\rightarrow 0$ in the limit $\textbf{g}_{tt}\rightarrow
0$, so that the photon suffers an infinite redshift. Thus a
surface defined by $\textbf{g}_{tt}=0$ is also often called an
infinite redshift surface. For the Schwarzschild spacetime, both
an infinite redshift surface and event horizon coincide, but in
more general axisymmetric spacetime such as Kerr black holes these
surfaces do not coincide. In our case, by setting
$-\textbf{g}_{tt}=B(r)=0$, we find that the surface of infinite
redshift is given by
\begin{eqnarray}\label{IRS2}
r_{s}=\left(\frac{3m}{\Lambda-b_0^2}\right)^{1/3}.
\end{eqnarray}
Thus the infinite redshift surface only exist provided we have
$\Lambda>b_0^2$. It can be seen that for $b_0=0$ the infinite
redshift surface coincides with the event horizon. This is indeed
an expected result, since in this case our solutions reduce to
black string solutions of Einstein gravity.
\begin{figure}[htp]
\begin{center}
\includegraphics[width=7cm]{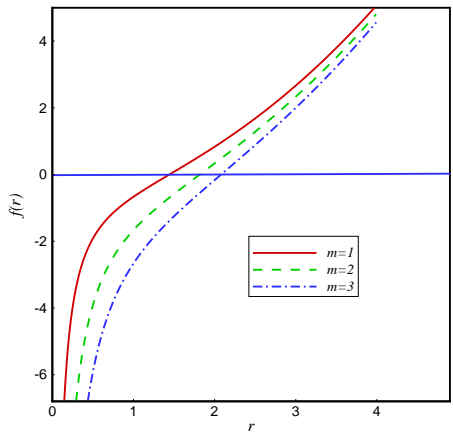}
\caption{The behavior of $f(r)$ for rotating black string with
$\Lambda=1$.}\label{Fig1}
\end{center}
\end{figure}
\begin{figure}[htp]
\begin{center}
\includegraphics[width=7cm]{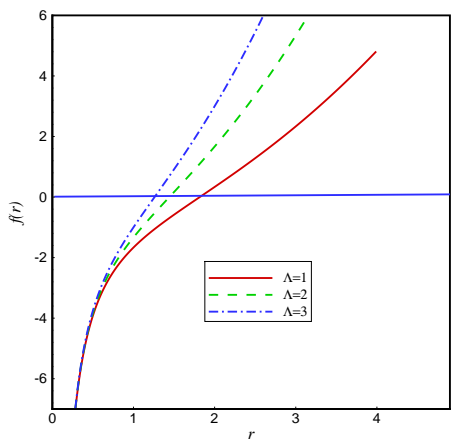}
\caption{The behavior of $f(r)$ for rotating black string with
$m=2$.}\label{Fig2}
\end{center}
\end{figure}
\begin{figure}[htp]
\begin{center}
\includegraphics[width=7cm]{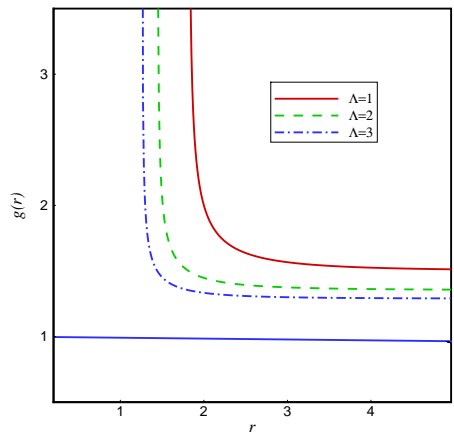}
\caption{The behavior of $g(r)$ for rotating black string. Here we
have taken $b_0=1/2$ and $m=2$.}\label{Fig3}
\end{center}
\end{figure}
The behaviour of the metric functions $f(r)$ and $g(r)$ are
depicted in Figs. 1-3. From Figs. 1 and 2 we see that rotating
black string in mimetic gravity has only one horizon at finite
radius $r_{+}$ where $f(r)$ and hence $B(r)=f(r)g^2(r)$ change
sign on it. We observe that with increasing $m$ or decreasing
$\Lambda$, the radius of this horizon increases as well, which can
be easily understood from its definition. Besides, as
$r\rightarrow0$, the metric function diverges, which confirm that
there is a singularity at $r=0$ which already confirmed that it is
indeed an essential singularity. Finally, from Fig. \ref{Fig3} it
is observed that the function $g(r)$ approach to constant value
$1+b_0/\sqrt{\Lambda}$ in large $r$ limit.

We finish this section by calculating the mass and angular
momentum of the rotating mimetic black strings. We shall use the
counterterm method inspired by AdS/CFT correspondence. In our case
the suitable boundary term which removes the divergences of the
action can be written
\begin{eqnarray}
S_{b} &=&\frac{1}{8\pi}\int_{\partial
\mathcal{M}}d^{3}x\sqrt{-\gamma } \left(\Theta(\gamma )+
\frac{2}{l}\right), \label{Actbound}
\end{eqnarray}
where $\Theta$ is the trace of the extrinsic curvature
$\Theta^{\mu \nu}$ for the boundary ${\partial \mathcal{M}}$ and
$\gamma$ is the induced metric on the boundary. The second term in
(\ref{Actbound}) is the the suitable counterterm. Variation of the
action (\ref{Actbound}) with respect to boundary metric
$\gamma_{\mu \nu}$ yields the following finite stress-energy
tensor \cite{mhd}
\begin{equation}
T^{ab}=\frac{1}{8\pi}\left[ \Theta^{\mu \nu}-\Theta\gamma ^{\mu
\nu}+\frac{2}{l}\gamma ^{\mu \nu}\right].  \label{Stres}
\end{equation}
In order to compute the conserved charges of the spacetime, we
choose a spacelike surface $ \mathcal{B}$ in the boundary
$\partial \mathcal{M}$ with metric $\sigma _{ij}$, and write the
boundary metric in ADM (Arnowitt-Deser-Misner) form:
\[
\gamma _{ab}dx^{a}dx^{a}=-N^{2}dt^{2}+\sigma _{ij}\left( d\varphi
^{i}+V^{i}dt\right) \left( d\varphi ^{j}+V^{j}dt\right) ,
\]
where the coordinates $\varphi ^{i}$ are the angular variables
parameterizing the hypersurface of constant $r$ around the origin,
and $N$ and $V^{i}$ are the lapse and shift functions
respectively. For the Killing vector field $\mathcal{\xi }$ on the
boundary, the quasilocal conserved quantities associated with the
stress tensors of Eq. (\ref{Stres}) are given by \cite{mhd}
\begin{equation}
Q(\mathcal{\xi )}=\int_{\mathcal{B}}d^{2}x \sqrt{\sigma }T_{ab}n^{a}%
\mathcal{\xi }^{b},  \label{charge}
\end{equation}
where $\sigma $ is the determinant of the metric $\sigma _{ij}$, $\mathcal{%
\xi }$ and $n^{a}$ are, respectively, the Killing vector field and
the unit normal vector on the boundary $\mathcal{B}$. For
boundaries with timelike ($\xi =\partial /\partial t$) and
rotational ($\varsigma =\partial /\partial \varphi $) Killing
vector fields, one obtains the conserved mass and angular momenta
of the system enclosed by the boundary $\mathcal{B}$ as \cite{mhd}
\begin{eqnarray}
M &=&\int_{\mathcal{B}}d^{2}x \sqrt{\sigma }T_{ab}n^{a}\xi ^{b},
\label{Mastot} \\
J &=&\int_{\mathcal{B}}d^{2}x \sqrt{\sigma }T_{ab}n^{a}\varsigma
^{b}.  \label{Angtot}
\end{eqnarray}
It is important to note that both $M$ and $J$ are independent of the particular choice of foliation $\mathcal{%
B}$ within the surface $\partial \mathcal{M}$ while they depend on
the location of the boundary $\mathcal{B}$ in the spacetime. Then,
it is a matter of calculations to show that the mass and angular
momentum per unit length of the rotating string when $
\mathcal{B}\rightarrow \infty$ are given by
\begin{eqnarray}\label{M}
{M}&=&\frac{(3\Xi ^{2}-1)m}{16\pi l}+\frac{b_0 m \sqrt{3}}{24\pi},
\end{eqnarray}
\begin{eqnarray}
{J}&=&\frac{3\Xi\sqrt{\Xi^2-1}m}{16\pi},  \label{J}
\end{eqnarray}
where we have also used $\Lambda=3/l^2$. When $a=0$ ($\Xi =1$),
the angular momentum per unit volume vanishes, which confirms that
parameter $a$ is indeed the rotational parameters of the black
string. Compared to the mass and angular momentum of black string
solutions of Einstein gravity \cite{mhd,Awad}, we see that the
angular momentum do not change,  while the mass get an additional
term. When $b_0=0$, the mass also reduces to the one of Einstein
gravity \cite{mhd,Awad}.
\section{Charged mimetic black string} \label{Cha}
Now we consider the general solutions given by (\ref{fr1}) and
(\ref{gr1}). Since we are looking for the analytical solutions and
the integral in (\ref{gr1}) cannot be done analytically, we study
the behaviour in two ranges. For large $r$, the dominant term in
the integrand is $\Lambda r^4$ and we have
\begin{eqnarray}\label{gr3}
g(r)&=&1+b\int{\frac{r^2 dr}{(|\Lambda r^4 -3m r+3q^2|)^{3/2}}}\\
&\approx &1+\frac{b}{\Lambda^{3/2}}\int{\frac{ dr}{r^4}
}+\cdot\cdot\cdot \approx
1-\frac{b}{3\Lambda^{3/2}}\frac{1}{r^3}+\cdot\cdot\cdot\label{gr3exp}
\end{eqnarray}
Thus the metric function $B(r)=f(r)g^2(r)$, for large $r$, is
\begin{eqnarray}\label{g002}
B(r)&=&-\frac{m}{r}+\frac{q^2}{r^2}+\frac{\Lambda
r^2}{3}-\frac{2b}{9\sqrt{\Lambda}} \frac{1}{r}+
O\left(\frac{1}{r^{4}}\right),
\end{eqnarray}
which contains an additional term of $O(1/r)$ compared to charged
rotating black string solution of Einstein gravity \cite{Awad}.
This term reflects the imprint of mimetic field on the spacetime.
We observe that the asymptotic behaviour of the solutions is AdS
which differs from the uncharged case where the spacetime was
approximately asymptotically AdS. On the other hand when
$r\rightarrow 0$, we can neglect the term $\Lambda r^4$ and the
solution becomes
\begin{eqnarray}\label{gr4}
g(r)&\approx&1+b\int{\frac{r^2 dr}{(3q^2-3mr)^{3/2}}}\approx
1+\sqrt{3} b
\left[2\left(\frac{2q}{3m}\right)^3+\left(\frac{r}{3q}\right)^3+\cdot\cdot\cdot\right],\label{gr3exp}
\end{eqnarray}
and thus
\begin{eqnarray}\label{g003}
B(r)&\approx&\left(-\frac{m}{r}+\frac{q^2}{r^2}+\frac{\Lambda
r^2}{3}\right)\Bigg{\{} 1+\sqrt{3} b
\left[2\left(\frac{2q}{3m}\right)^3+\left(\frac{r}{3q}\right)^3+\cdot\cdot\cdot\right]\Bigg{\}}^{2}.
\end{eqnarray}
From (\ref{g003}) we observe that near the singularity where
$r\rightarrow 0$ we have $B(r)\rightarrow\infty$, as one expected.
When $b=0$, our solutions given in (\ref{g002}) and (\ref{g003})
reduce to asymptotically AdS charged black string of Einstein
gravity.

The horizon can be obtained by solving $f(r)=0$, where $f(r)$ is
given by Eq. (\ref{fr1}). Thus, the horizons are the real root of
equation $\Lambda r^4 -3m r+3q^2=0$. Depending on the values of
the parameters this equation may have zero, one or two roots. The
cases with zero or one root correspond to naked singularity and
extremal black hole, respectively. When this equation has two
roots we encounter a black hole with two inner and outer horizon.
\begin{figure}[htp]
\begin{center}
\includegraphics[width=7cm]{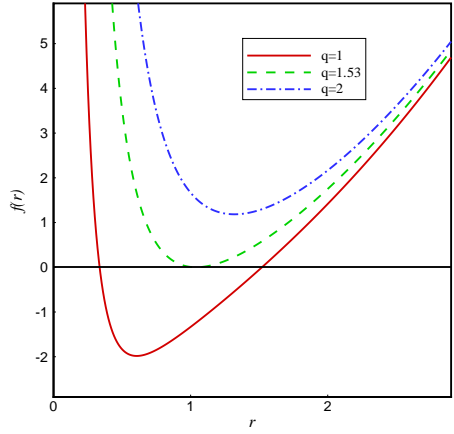}
\caption{The behavior of $f(r)$ for charged rotating black string
with different $q$, where we have taken $m=3$,
$\Lambda=2$.}\label{Fig6}
\end{center}
\end{figure}
\begin{figure}[htp]
\begin{center}
\includegraphics[width=7cm]{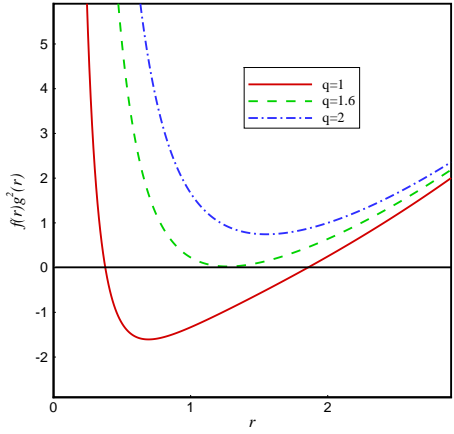}
\caption{The behavior of $B(r)=f(r)g^2(r)$ for charged rotating
black string given in Eq. (\ref{g002}) with different $q$. Here we
have taken $m=2$, $b=3$ and $\Lambda=1$.}\label{Fig7}
\end{center}
\end{figure}

\begin{figure}[htp]
\begin{center}
\includegraphics[width=7cm]{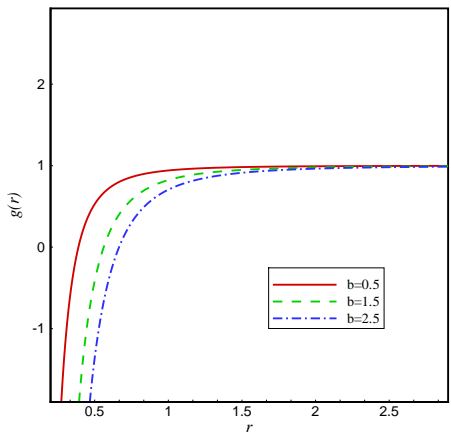}
\caption{The behavior of $g(r)$ for charged rotating black string
for large $r$ limit with $\Lambda=2$.}\label{Fig8}
\end{center}
\end{figure}

\begin{figure}[htp]
\begin{center}
\includegraphics[width=7cm]{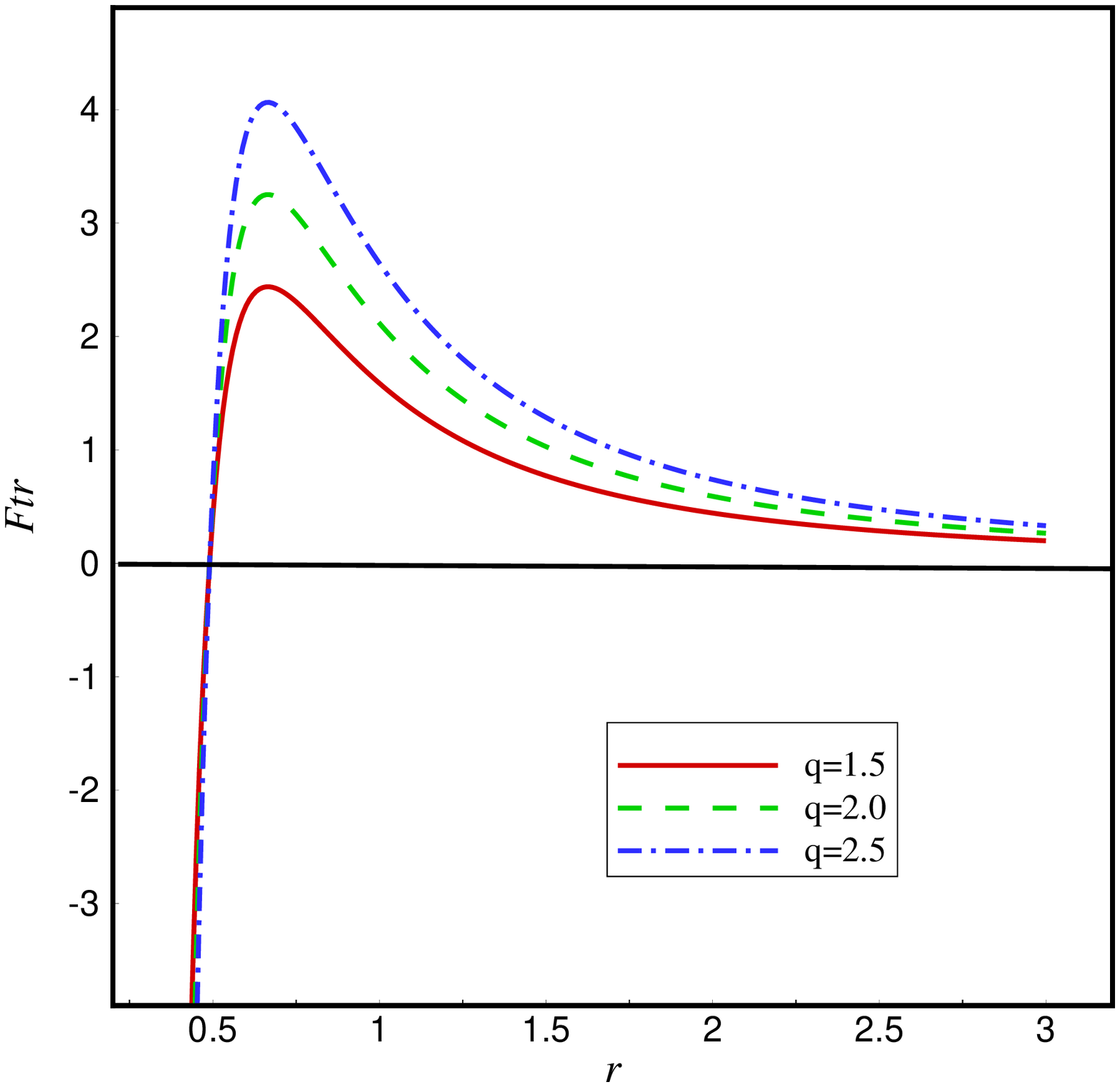}
\caption{The behavior of $F_{tr}(r)$ for charged rotating black
string for different $q$. Here we have taken $b=1$, $\Lambda=2$,
$\Xi=1.2$.}\label{Fig9}
\end{center}
\end{figure}

\begin{figure}[htp]
\begin{center}
\includegraphics[width=7cm]{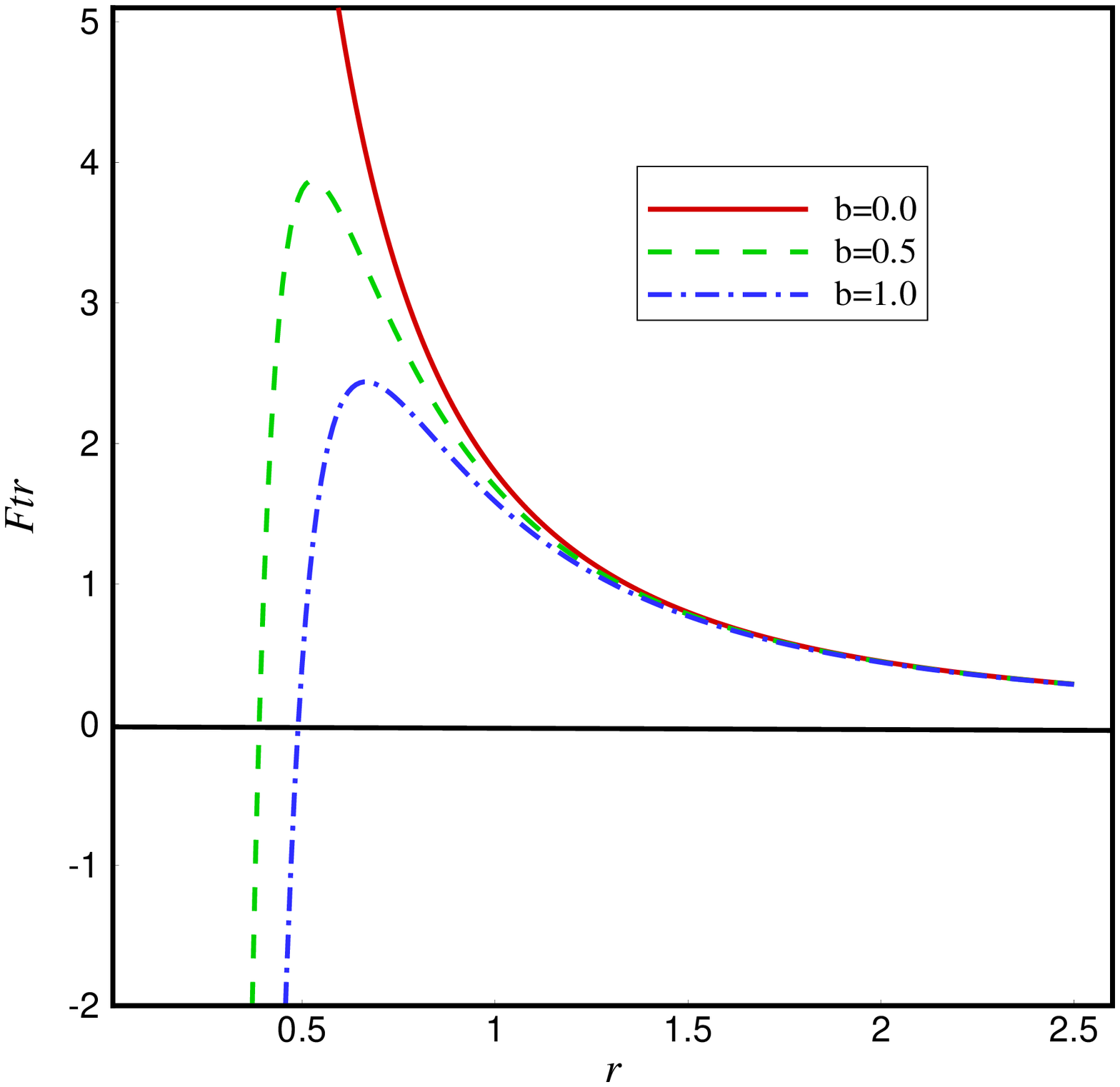}
\caption{The behavior of $F_{tr}(r)$ for charged rotating black
string for different parameter $b$. Here we have taken $q=1.5$,
$\Lambda=2$, $\Xi=1.2$.}\label{Fig10}
\end{center}
\end{figure}
The behavior of the metric functions for charged rotating black
string are depicted in Figs. \ref{Fig6}-\ref{Fig8}. From Fig.
\ref{Fig6}, we see that, depending on the metric parameters, our
solutions can represent black string with two horizons, an
extremal black string or naked singularity. From Fig. \ref{Fig7},
we also observe that our solutions admit two, one or zero surface
of infinite redshift if the metric parameters are chosen suitably.
We have also plotted the bahviour of $g(r)$ in Fig. \ref{Fig8},
where it can be seen that $g(r)\rightarrow 1$ for large values of
$r$ and  goes to $-\infty$ as $r\rightarrow0$.

Given the function $g(r)$, the explicit form of the electric
field, for large $r$, is given by
\begin{equation}\label{Ftrexp}
F_{tr}=\frac{q \Xi }{r^2}\left[
1-\frac{b}{3\Lambda^{3/2}}\frac{1}{r^3}+
O\left(\frac{1}{r^5}\right)\right].
\end{equation}
Let us note that for $b\neq0$, the electric field has a maximum
value at $r_{\rm max}= \left(\frac{5b}{6
\Lambda^{3/2}}\right)^{1/3}$, and goes to zero as $r\rightarrow
\infty$. This maximum value appears since the imprint of the
mimetic field which contributes to the second negative term of
(\ref{Ftrexp}) reduces the electric field compared to the case of
Einstein gravity. Figs. \ref{Fig9} and \ref{Fig10} demonstrate the
behaviour of the electric field $F_{tr}(r)$ for charged rotating
mimetic black strings. From these figures we observe that the
maximum value of electric field increases with increasing $q$,
while it decreases and shifts to larger $r$ with increasing $b$.
This can be understood since for larger $b$ the effects of the
mimetic field makes the electric field weaker. The large $r$ limit
behaviour of the metric functions as well as $F_{tr}(r)$ and
$F_{\varphi r}(r)$ indicate that far from the black string $(r\gg
r_{+})$, the effects of the mimetic field disappear and one
recovers the Einstein gravity up to a constant.

We then look for the curvature singularity. To do that, we expand
the scalar invariants for the obtained solutions,
\begin{eqnarray}
&&\lim_{r\rightarrow0}R =-\frac{126 q^2} {r^{4}}+\cdot\cdot\cdot  \label{Rorigin2} \\
&&\lim_{r\rightarrow 0}R_{\mu \nu \rho \sigma }R^{\mu \nu \rho
\sigma }=\frac{26124 q^4}{r^8}+\cdot\cdot\cdot \label{RRorigin2}
\end{eqnarray}
Therefore, we have an essential singularity located at $r=0$. On
the other side, for the asymptotic region where $r\rightarrow
\infty$, the invariants of the spacetime are obtained as
\begin{eqnarray}
 &&\lim_{r\rightarrow \infty }R =-4\Lambda ,  \label{Rinf} \\
&&\lim_{r\rightarrow \infty }R_{\mu \nu \rho \sigma }R^{\mu \nu
\rho \sigma } =\frac{8}{3}\Lambda ^{2}. \label{RRinf}
\end{eqnarray}
In order to calculate the mass and angular momentum of charged
rotation black string, we use the method of the previous section.
In this case the mass and angular momentum per unit length of the
charged rotating black string are given by
\begin{eqnarray}\label{M2}
{M}&=&\frac{(3\Xi ^{2}-1)m}{16\pi l}+\frac{b
(\Xi^2-1)\sqrt{3}}{72\pi},
\end{eqnarray}
\begin{eqnarray}
{J}&=&\frac{\Xi \sqrt{\Xi^2-1}(27m+2 \sqrt{3}b l)}{144\pi}.
\label{J2}
\end{eqnarray}
When $b=0$, the above expressions for the mass and angular
momentum restore their respective expressions in Einstein gravity.
Let us note that $M$ and $J$ differ for charged and uncharged
rotating black strings. This may due to the fact that the
asymptotic behaviour of these two classes of solutions are
different. For the uncharged string, the asymptotic behaviour is
\textit{approximately} AdS, while for charged rotating string, the
solutions are asymptotically AdS. However, this issue deserves
further investigation.

Finally, we calculate the electric charge of the mimetic black
string. For this purpose, we first determine the electric field by
considering the projections of the electromagnetic field tensor on
special hypersurface. The normal vectors to such hypersurface are
\begin{equation}
u^{0}=\frac{1}{N},\text{ \ }u^{r}=0,\text{ \
}u^{i}=-\frac{V^{i}}{N},
\end{equation}
where $N$ and $V^{i}$ are the lapse function and shift vector.
Then the electric field is $E^{\mu }=g^{\mu \rho }F_{\rho \nu
}u^{\nu }$, and the electric charge per unit length of the string
can be found by calculating the flux of the electric field at
infinity, yielding
\begin{equation}
{Q}=\frac{\Xi q}{4\pi l}.  \label{Q}
\end{equation}
\section{Conclusion and discussion}\label{Con}
In conclusions, we have obtained two new classes of rotating black
string solutions in the context of mimetic gravity and in the
presence of a constant potential, $V(\phi)=-2\Lambda$ for the
mimetic field $\phi$. We first, derived  exact analytical
solutions for uncharged rotating black string and investigated
their physical properties. In this case, we found that the imprint
of the mimetic field on the gravitational field equations makes a
deviation from the AdS space for the the asymptotic behvaiour of
the solution, although one can still has AdS solution by
redefinition of $\Lambda$. We also obtained the surface of
infinite red shift and observed that it is different from the
event horizon and only exist for $\Lambda>b_0^2$. When $b_0=0$,
the infinite redshift surface coincides with the event horizon.
Then, we explored charged rotating mimetic black string. In this
case we could obtain analytical solutions, after expanding the
integrand in Eq. (\ref{gr1}), namely for large/small values of
$r$. In this case our solution is asymptotically AdS and may have
no horizon (naked singularity), one or two horizons depending on
the metric parameters. Besides, both electric field $F_{tr}$ and
magnetic field $F_{\varphi r}$ diverge for small $r$, they have a
maximum value at finite $r$ and go to zero as $r\rightarrow
\infty$.

For both class of solutions, we also observed that the Kretschmann
invariant $R_{\mu \nu \rho \sigma }R^{\mu \nu \rho \sigma }$ and
the Ricci scalar diverge at $r=0$, and they are finite for $r\neq
0$. Thus, there is a n essential singularity located at $r=0$.
Besides, $R_{\mu \nu \rho \sigma }R^{\mu \nu \rho \sigma }
\rightarrow8 \Lambda^2/3 $ and $R\rightarrow -4\Lambda$ as
$r\rightarrow \infty$. In the absence of the mimetic field, our
solutions reduce to the asymptotically AdS rotating black string
in Einstein gravity \cite{mhd,Awad}. We also calculated the mass
and angular momentum per unit length of the rotating black strings
which has an additional term compared to the case of rotating
string in Einstein gravity \cite{mhd,Awad}.

Let us confess that we also made some attempts to construct exact
analytical solutions in case of variable potential, e.g.,
$V[\phi(r)]=V(r)$. Unfortunately, we failed to find \textit{exact}
analytical solutions which fully satisfy the field equations
\footnote{Note that the obtained solutions in Ref. \cite{Myr2} are
not exact analytical solutions, and only valid for some ranges of
$r$. Indeed they constructed potential for the mimetic field, by
using an inverse approach, i.e., by guessing a metric function and
inserting it into the field equations to derive the functional
form of the potential. Their solutions are not the general
solutions of the full field equations with variable potential and
they take some approximations or neglecting some terms. Here, we
tried to insert a variable potential, as an input, in the field
equations and derive analytically the corresponding exact metric
functions}. For instance, if we examine $V=-2\alpha r^{-n} $ we
obtain, for uncharged case ($q=0$), the following solution for Eq.
(\ref{ppzz})
\begin{eqnarray}
f(r)=\frac{\alpha r^{2-n}}{3-n}-\frac{c_1}{r}.
\end{eqnarray}
Then depending on the value of $n$, one can solve the field
equations (\ref{tt})-(\ref{pt-tp}) to obtain the function $g(r)$.
For instance, for $n=2$ these equations admit the following
consistent unique solution
\begin{eqnarray}
g(r)=c_1+2\alpha r +\frac{c_2 r^{3/2}}{\sqrt{\alpha r-c_1}},
\end{eqnarray}
where $c_1$ and $c_2$ are integration constants \footnote{For
$n=-2$ and $n=\pm 1$, the function $g(r)$ is in the form of
hypergeometric function, while for $n=\pm 1/2$, it is in the form
of Legendre function}. The main challenge is that the obtained
solutions for variable potential do not satisfy the equation of
motion (\ref{FEE2}) of the scalar field. We have examined a wide
range of functions for the potential and arrived at the same
conclusion; the obtained solutions from the Einstein equations
(\ref{FEE1}) do not satisfy the equation of motion (\ref{FEE2})
for the scalar field. This may be due to the nature of mimetic
gravity which implies that $\phi$ is not a dynamical scalar field
by itself, but only makes the longitudinal mode of gravity to be
dynamical. However, this issue is not well settle down and
certainly needs further clarification.

Many issues are remained for future investigations. First of all,
this study can be generalized to higher dimensional spacetime by
exploring the rotating black branes/holes in the framework of
mimetic gravity. One may also try to derive $(2+1)$-dimensional
BTZ like solutions of mimetic gravity. Investigation the geodesic
motion of massless and massive particles around this spacetime is
also another task. It is also interesting to explore
thermodynamics, thermal stability, and phase transition of mimetic
black holes/strings. We leave these issue for future studies.


\acknowledgments{I am grateful to Shiraz University Research
Council. I also thank Max-Planck-Institute for Gravitational
Physics (AEI), where this work written and completed, for
hospitality.}


\end{document}